# Particle Physics at Accelerators in the United States and Asia

**Pushpalatha C. Bhat[1,‡] and Geoffrey N. Taylor[2]**

[1]*Fermi National Accelerator Laboratory*
*Batavia, IL, USA*

[2]*School of Physics*
*University of Melbourne*
*Melbourne, Australia*

**Abstract:**

**Particle physics experiments in the United States and Asia have greatly contributed to the understanding of elementary particles and their interactions. With the recent discovery of the Higgs boson at CERN, interest in the development of next-generation colliders has been rekindled. A linear electron-positron collider in Japan and a circular collider in China have been proposed for precision studies of the Higgs boson. In addition to the Higgs programme, new accelerator-based long-baseline neutrino mega-facilities are being built in the United States and Japan. Here, we outline the present status of key particle physics programmes at accelerators and future plans in the United States and Asia that largely complement approaches being explored in the European Strategy for Particle Physics Update. We encourage the pursuit of this global approach, reaching beyond regional boundaries for optimized development and operations of major accelerator facilities worldwide, to ensure active and productive future of the field.**

The discovery [1,2] of the Higgs boson in 2012 completed the Standard Model (SM) of particle physics while also opening a new portal to the unknown. In the years since, the data samples accumulated by the ATLAS and CMS experiments at the Large Hadron Collider (LHC) have grown 25-fold allowing for measurements of a range of properties of the Higgs boson and its couplings to many SM particles to 10-25% precision [3,4]. No evidence of any excursion from SM predictions has been found, but the hopes that the "Higgs portal" can be exploited to search for phenomena beyond the SM persist. The collision energy of the LHC is to be raised to 14 TeV in the next data-taking period scheduled for 2021–2023, aiming to double the data set for a total integrated luminosity of ~300 $fb^{-1}$.

The high-luminosity upgrade of the LHC will increase the proton–proton collision data sets by an order of magnitude to 3000 $fb^{-1}$ over the operation period 2027–2037. Most Higgs couplings will then be measurable to a few percent accuracy [5], allowing either discovery of deviations from SM predictions, — potentially evidence for new physics — or further show the SM to be accurate to an astonishing precision.

The Higgs discovery has renewed interest worldwide in the development of next generation colliders. Due to the Higgs mass of 125 GeV, several proposals for an electron–positron Higgs Factory have been put forth: the 250 GeV International Linear Collider (ILC) [6-8] in Japan, the 380 GeV Compact Linear Collider (CLIC) [9] based on a novel two-beam acceleration technique at CERN, the Future Circular Collider [10] (FCC-ee) at CERN, and the Circular Electron–Positron Collider (CEPC) [11] in China. Both FCC-ee and CEPC are proposed to be followed by a proton–proton collider. Such an electron–positron machine is expected to provide exquisitely precise measurements of the Higgs boson couplings, in search of deviations from the SM. At higher energies, direct measurements of the Higgs self-coupling, essential to unravel the details of electroweak symmetry breaking, would also be possible.

A lepton collider would complement the broad exploration as well as precision SM measurements at the LHC. It would also be complementary to the ongoing and planned worldwide programmes in neutrino physics, direct dark matter detection, gravitational wave measurements, and the projects in astrophysics aimed at the study of dark energy. This strong and diverse portfolio of scientific projects would be unprecedented in the history of physics and may yield discoveries and insights that could transform our understanding of the Universe.

---
[‡] Email: pushpa@fnal.gov

**Past and Present Particle Physics Programmes in the United States and Asia**

Particle physics is a vibrant field in the United States and Asia. The United States has a rich history in particle accelerators and major discoveries in particle physics made at US high energy accelerator facilities (for example, at Brookhaven National Laboratory, Stanford Linear Accelerator Center (SLAC) and Fermi National Accelerator Laboratory (Fermilab)): the muon neutrino [12], evidence for the quark model [13,14], the charm quark [15,16], the tau lepton [17,18], the bottom quark [19], the top quark [20,21] and the tau neutrino [22]. Fermilab's Tevatron, which operated for well over 25 years, had been the world's most powerful proton/anti-proton accelerator and collider prior to the commissioning of the LHC at CERN. In Japan and China, electron–positron colliders as well as rich neutrino physics programmes have led to results of central importance to particle physics (as discussed later in this section).

Particle physics has seen important, constructive transatlantic competition and cooperation between the United States and Europe for several decades with parallel and lock-step developments of proton synchrotrons, electron linear accelerators and colliders. There were also setbacks, the foremost being the termination of the Superconducting Super Collider in the United States in 1993. Nonetheless, the success of the field continues to owe much to the combination of competitive urge and cooperative developments.

In Asia, Japan's National Laboratory for High Energy Physics (KEK) built the electron–positron collider TRISTAN following the 8–12 GeV proton synchrotron, whilst in China the Beijing Electron–Positron Collider (BEPC), upgraded to BEPC-II at the Institute of High Energy Physics has evolved into a world-leading programme in tau-lepton and charm-quark physics [23]. Examples of trans-pacific competition emerged between the Babar (b and b-bar) experiment at SLAC in the United States and the Belle experiment at KEK in Japan, as well as with BESII (Beijing Spectrometer II) at BEPC in China.

Additionally, the United States, China, Japan, Korea, India, Taiwan, Australia and others are active participants in international collaborations. In the era of the LHC, partnerships with CERN have significantly grown and strengthened. The United States has the single largest national contingent of physicists at CERN making important contributions to all major LHC collaborations and the accelerator projects. CERN will continue to be a focus of United States and Asian physics programmes into the foreseeable future.

The KEKB facility in Japan has now been upgraded to the SuperKEKB collider aiming for a leap of a 40 times the luminosity. The improved Belle II detector is sensitive to new physics in the flavour sector, similar to the LHCb experiment at the LHC; the two have highly complementary capabilities to different regions of phase space and decay modes, and could find evidence for new physics if the current flavour anomalies persist [24].

The United States, Japan, China and Korea have pursued studies of neutrinos, producing very important results. The Japan Proton Accelerator Research Complex (J-PARC) with a 30 GeV proton synchrotron enabled a successful long-baseline neutrino programme from Tokai to Kamioka (T2K) after the KEK to Kamioka (K2K) experiment verified neutrino oscillations that were first observed with atmospheric neutrinos using the Super-Kamiokande detector [25]. The Sudbury Neutrino Observatory (SNO) Collaboration also verified the neutrino oscillation hypothesis by detecting all flavours of neutrinos using a heavy water detector and showing that about two-thirds of the electron neutrinos from the Sun had changed to other flavours [26] on their way to Earth. China's Daya Bay [27] and Korea's Reactor Experiment for Neutrino Oscillation (RENO) [28] nuclear reactor neutrino experiments have impacted our understanding of neutrino oscillations significantly. Daya Bay has measured the neutrino mixing angle $\theta_{13}$ to world-leading precision of 4%, opening up the search for charge-conjugation/parity symmetry violation (CP violation) in the lepton sector. The NuMI Off-axis $\nu_e$ Appearance (NOvA) experiment making use of Fermilab's Main Injector neutrino complex, currently operating with up to 750 kW proton beam on target (undergoing further upgrades), and a far detector 800 km away in Minnesota, will measure not only neutrino oscillation parameters, but will discern the hierarchy of neutrino masses and possibly observe CP violation. The NOvA and T2K collaborations are working together to combine their results for enhanced significance. An international short baseline neutrino oscillation programme starting with Fermilab's 8 GeV Booster beam is also flourishing; MicroBooNE is taking data, ICARUS and the Short Baseline Near Detector (SBND) will soon be underway. Primary science goals include search for new physics, particularly for sterile neutrinos, and study of neutrino–nucleus interactions, which help reduce systematic uncertainties in measurements of neutrino oscillation parameters.

Small and medium scale precision physics measurements in both the United States and Asia search for new particles and interactions in an indirect way. To measure the gyromagnetic factor g, or rather "g-2", of the muon to a precision of 140 parts per billion, the Muon g-2 experiment is underway at Fermilab [29]. This promises a four-fold improvement over the previous measurement made at Brookhaven National Laboratory [30], which was discrepant from the SM expectation by 3.5 standard deviations. The Mu2e experiment at Fermilab and the COMET (Coherent Muon to Electron Transition) experiment at J-PARC in Japan will look for neutrino-less muon-to-electron conversion in the Coulomb field of the nucleus, which — if observed — signals charged lepton flavour violation and would be evidence for new physics beyond the SM. The sensitivity goals of these measurements is aimed at reaching levels four orders of magnitude better than the previous experiments.

The BEPC II electron–positron collider in Beijing and its BESIII detector, continue with productive low-energy experiments with a series of discoveries and measurements of the so-called X, Y, and Z resonances ([31] and references therein). Both China and Russia have plans for high-intensity electron–positron machines serving as tau-lepton and charm-quark factories in the 3 GeV region.

**The Future of Particle Physics in the United States and Asia**

The U.S. and Asian particle physics activities in the upcoming decade focus mainly on neutrino physics and the exploration of the Higgs sector. Participation in the broader physics programme at the energy frontier at CERN will continue.

*Neutrino Physics*

The United States is building its next flagship neutrino facility — the Long Baseline Neutrino Facility (LBNF) to host the Deep Underground Neutrino Experiment (DUNE) [32], to be completed in the late 2020s. Proton beam power will be upgraded into the MW regime. Construction of the 800 MeV Super Conducting Radio Frequency (SCRF) PIP-II (Proton Improvement Plan-II) linear accelerator, is underway. Ultra-intense neutrino beams will travel from Fermilab some 1300 km through the Earth's mantle, as shown in Fig. 1, to a massive 70 kton Liquid Argon detector assembly (with four 10 kton Liquid Argon detectors) located 1.6 km beneath the surface. A versatile near detector at the Fermilab site will enable detailed studies of neutrino interactions and control systematic uncertainties. The DUNE experiment will allow for a definitive measurement of the neutrino mass hierarchy, observation of leptonic CP violation over a large parameter space and precision measurement of mixing parameters. With the DUNE detector, proton decay, atmospheric neutrinos, and neutrinos of astronomical origin can also be studied.

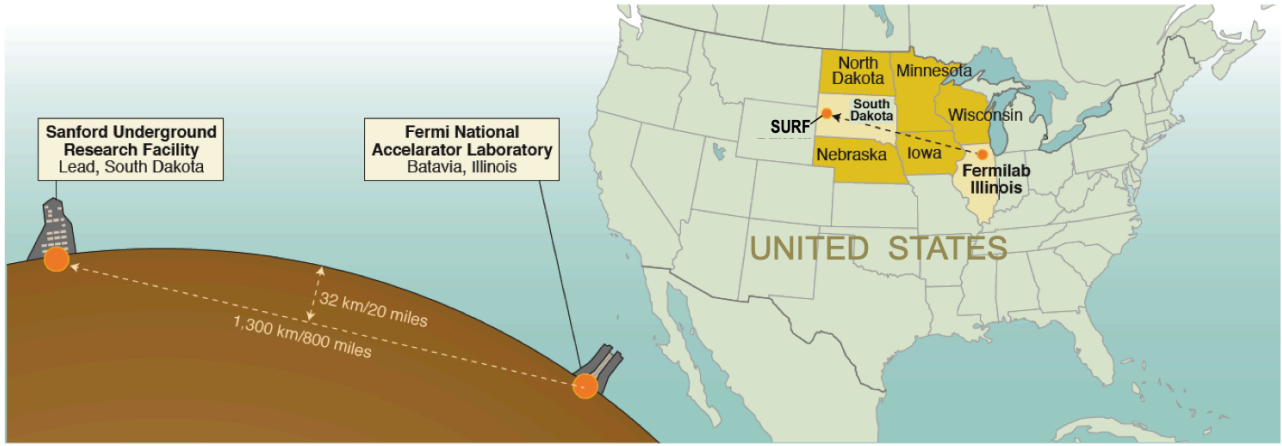

**Fig. 1. Schematic of LBNF/DUNE.** Schematic of the Fermilab-based Long Baseline Neutrino Facility (LBNF) for the Deep Underground Neutrino Experiment (DUNE). Fermilab plans to produce and send the world's most powerful neutrino beam to the DUNE experiment to be installed in a mine 1300 km away and 1.6 km below surface at the Sanford Lab in South Dakota. (Credit: Fermilab)

LBNF/DUNE is the first such international mega-project in the United States; the DUNE collaboration is modelled after ATLAS and CMS, with contributions to come from 31 countries. Partnership with CERN in the prototype construction — the proto-DUNE detectors, and their testing on the dedicated Neutrino Platform — is shaping to be pivotal to the eventual successful execution of DUNE. Also, PIP-II is the first accelerator to be built in the United States with five countries making substantial contributions.

Japan is continuing its successful accelerator-based long baseline neutrino oscillation program with the Hyper-Kamiokande (Hyper-K) project comprising the Hyper-Kamiokande detector, and upgrades to the J-PARC accelerator complex. The latter has commenced development, having achieved proton beam power greater than 500 kW. The Hyper-K project has recently received initial funding and project approval from the Japanese government. The potential of Hyper-K is similar to LBNF/DUNE, but uses different detector technology (water Cherenkov detectors) successfully employed in its predecessors. The key parameter of L/E is similar for both, but different values of L (baseline length; distance from neutrino production point to the detector) and E (neutrino energy) give independent sensitivities to the two experiments for mass hierarchy and CP violation.

*Higgs Physics and the Energy Frontier*

While the U.S. and Asian groups continue to contribute to the LHC and its high-luminosity upgrade, there is consensus in the global physics community that the next collider facility for particle physics should be an electron–positron Higgs factory. The Higgs boson can be copiously produced in electron–positron collisions at 250 GeV via the process $e^+e^- \to Z^0H$ (see Fig. 2). The backgrounds to this signature, primarily from radiative $e^+e^- \to Z\gamma$ and $e^+e^- \to Z^0Z^0$ processes, are well understood and can be computed from electroweak theory at the 0.1% level [33]. The very clean signal process with low background has the distinct advantage of Higgs identification via the recoiling $Z^0$, allowing for an unbiased selection of H decays — even invisible modes. Measurements of Higgs couplings to SM particles with percent-level precision or better should be feasible, providing sensitivity to physics beyond the SM. Appropriate polarisation of both beams can enhance the Higgs production cross sections as well as provide a powerful tool to probe new physics. Potential upgrades to 350 GeV and beyond would allow precise measurements of the top quark mass and coupling, as well as access to Higgs self-coupling.

The required collision energy of only 250 GeV for the Higgs factory opens up various options for the collider technology. Linear colliders have been in development for the past few decades. The ILC design and technology is the most mature and ready for construction. The critical technology driving both performance and price is the SCRF cavity that has seen major improvements in recent years. Furthermore, the European XFEL (X-ray Free Electron Laser) at DESY (Deutsches Elektronen-Synchrotron), built with 101 cryogenic modules with SCRF cavities providing 23.6 MV/m accelerating gradient, and delivering a 17.5 GeV electron beam, can be considered a prototype of this key technology.

Japan's Ministry of Education, Culture, Sports, Science, Technology (MEXT) has been considering hosting the 250 GeV ILC in Japan with international partnerships. Recent R&D developments via a U.S.–Japan collaboration include a significant increase in the acceleration gradient of SCRF cavities, and reduced construction costs. Ability to polarise the beams, the possibility of future energy upgrades, and a reduced price-tag have put the ILC in the spotlight for funding. With a machine cost of the same order as that of the LHC, the ILC is deemed to be within reach of international resources and capabilities.

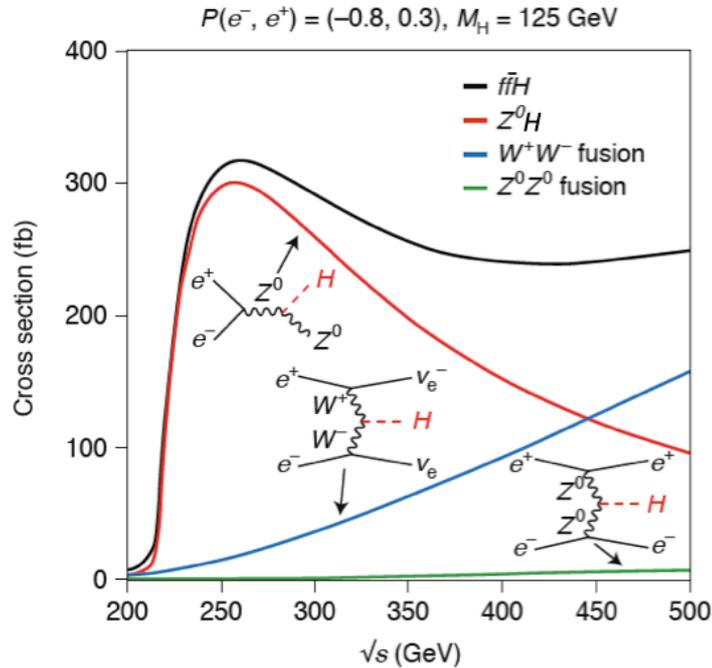

**Fig. 2: Dominant Higgs production cross sections.** Higgs production cross sections as a function of centre-of mass-energy for three major processes and the corresponding Feynman diagrams. Beam polarization of -0.8 for electrons and +0.3 for positrons, and Higgs mass of 125 GeV are assumed. The dominant production process is the Higgsstrahlung process where the Higgs boson is produced in association with a $Z^0$ boson (red curve), followed by the two vector boson fusion processes, $W^+W^-$ fusion process (blue curve) and the $Z^0Z^0$ fusion process (green curve). The black curve shows the total cross section for Higgs production with fermion pairs. (Adapted from Ref. [33])

Closely following the ILC is the innovative CLIC design with beam-driven acceleration technology, capable of considerably higher energy. The CLIC proposal offers a staged approach with an initial collision energy of 380 GeV to allow both the Higgs factory mode at 250 GeV and also to better explore the physics of the top quark. Linear colliders offer the benefit of potential energy upgrades without the limitation from synchrotron radiation of circular machines. The outcome of deliberations in the European Strategy for Particle Physics Update will impact upon expectations for a linear collider in Europe or Japan.

The International Committee for Future Accelerators (ICFA) [34,35], created to facilitate strong international partnerships in all aspects of construction and exploitation of large accelerator facilities, has actively engaged in efforts to enable the realization of a linear electron–positron collider since the 2000s. ICFA set up the International Linear Collider Steering Committee in 2002, and the Global Design Effort in 2005 that led to the ILC Technical Design Report [6] in 2013. The Linear Collider Board, set up by ICFA, has been pivotal in advancing the physics case of a linear collider Higgs Factory, coordinating global efforts on the accelerator and detector technology R&D and communicating with the particle physics community.

In March 2019 at the ICFA/Linear Collider Board meeting in Tokyo, Japan's MEXT acknowledged the importance of the ILC and requested KEK to initiate an international working group to consider various aspects of the project. The group has reported on proposed international cost sharing and possible management schemes [36] for an ILC laboratory. Preliminary discussions about possible partnership arrangements have commenced between MEXT/KEK and representatives in the United States, France and Germany. Formal negotiations await a positive commitment by the Japanese government to pursue hosting the ILC.

Circular electron–positron colliders such as the FCC-ee in a 100 km tunnel can operate at very high luminosity from the $Z^0$ boson to the top-pair production threshold. A parallel proposal in China to build the CEPC, a 100km circular collider initially operating at collision energies 240-250 GeV, is also reaching maturity [11].

While lepton colliders are critical for precision measurements, much less demanding and could provide access to new physics, hadron colliders are essential for a broad exploration of the high-energy frontier, and hence the two are complementary. The significant increase in cross-section with energy is a strong driver for proposing higher energy, post-LHC, proton–proton colliders. The proposed proton–proton facility, FCC-hh, to follow the FCC-ee can also be augmented to study electron–proton collisions and ion–ion collisions. As in the case of the FCC, China proposes to use the CEPC tunnel to install the Super Proton–Proton Collider (SPPC) after the CEPC programme.

The combination of a very large circular electron–positron collider — with the potential future proton–proton collider — in parallel with a 250 GeV (and higher) energy electron–positron linear collider would provide a comprehensive, powerful and complementary set of facilities for the world's particle physics programme. Whether the cost of such a set of machines can be met is yet to be determined. With prospects for the ILC in Japan and CEPC in China, there is the potential for Asia to emerge in the coming years as a leading world region for the field.

**Shared Technology Development**

As the field embarks on designing and building frontier facilities with strong international partnerships, the development of technologies are carried out and shared globally. Highly-developed, industrial-scale accelerator dipole magnets have been critical for pushing the high-energy circular proton–proton colliders. Rapidly developing high field-gradient radio frequency cavities for acceleration, displayed in Fig. 3, are critical for future facilities, especially for linear colliders.

To facilitate realization of a 250 GeV ILC (schematic shown in Fig. 4), the Linear Collider Collaboration, overseen by ICFA's Linear Collider Board, has been coordinating worldwide efforts on SCRF R&D, and other accelerator technology challenges. There is ongoing U.S.–Japan collaborative work on cost reduction via SCRF cavity performance improvements. Significant progress has been made in both the gradient and the quality factor of the cavities. Gradients in excess of 45 MV/m have been achieved [37,38], exceeding the original ILC design of 31.5 MV/m. Researchers continue to push the limits of this technology, which offers greater energy in a fixed length facility. With prospects for reaching gradients as high as 90 MV/m, such cavities could potentially triple the available collision energy within the proposed ~20 km ILC tunnel. KEK and CERN are collaborating on addressing several other technical challenges essential for the ILC, including nano-beam production and high-power beam dump designs.

Asian laboratories are working with U.S. labs in collaboration with CERN on high-field magnet development to enable future proposed 100 km proton colliders to operate at energies in excess of 100 TeV. China has commenced R&D into iron-based superconductors, which may offer desirable mechanical properties and higher fields at a cheaper cost. Similar efforts are also beginning in the United States.

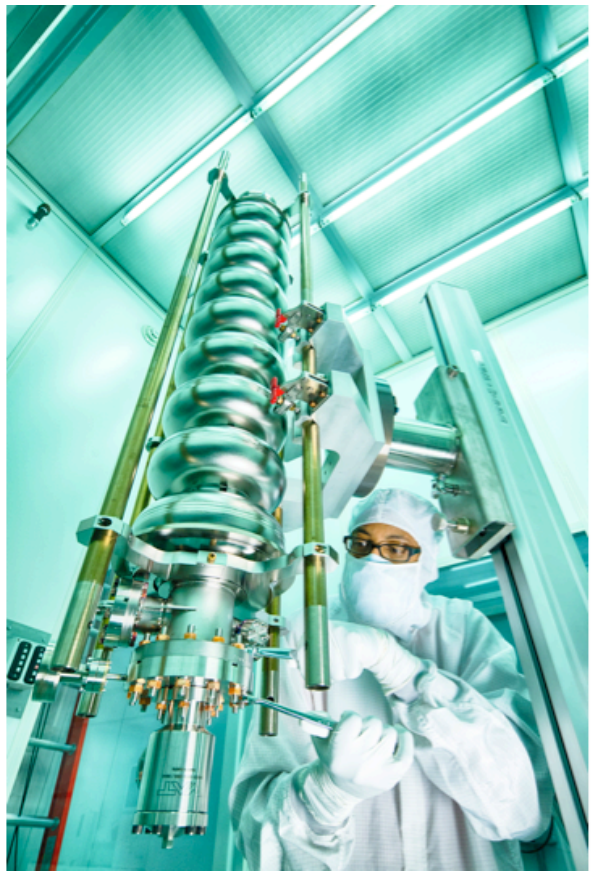

**Figure 3. Superconducting radio-frequency cavity.** Superconducting radio-frequency (SCRF) cavities will be the primary components of the International Linear Collider. Here, an SRF nine-cell cavity is being worked on at the cavity processing facility at Fermilab. (Credit: Fermilab)

Muon colliders continue to be of interest [39] with the potential to reach collision energies well over 10 TeV, but their development has many challenges including cooling of the muons. The international MICE (Muon Ionization Cooling Experiment) collaboration has recently demonstrated muon cooling [40]. An innovative approach of producing low-emittance $\mu^+\mu^-$ pairs [41] followed by acceleration and high-energy collisions is also under study. The LHC tunnel may be well suited for a future facility hosting a 14 TeV muon collider [42].

The promise of very high accelerating gradients in plasma-based accelerators [43-45] is enticing. Despite the success of proof-of-principle experiments [46,47], there are formidable challenges in realizing a high-energy collider. The ICFA panel on Advanced and Novel Accelerators has begun to facilitate joint studies and R&D plans for such accelerators with potential applications and with a view of developing a design for an advanced linear electron–positron collider by the mid-2030s [48].

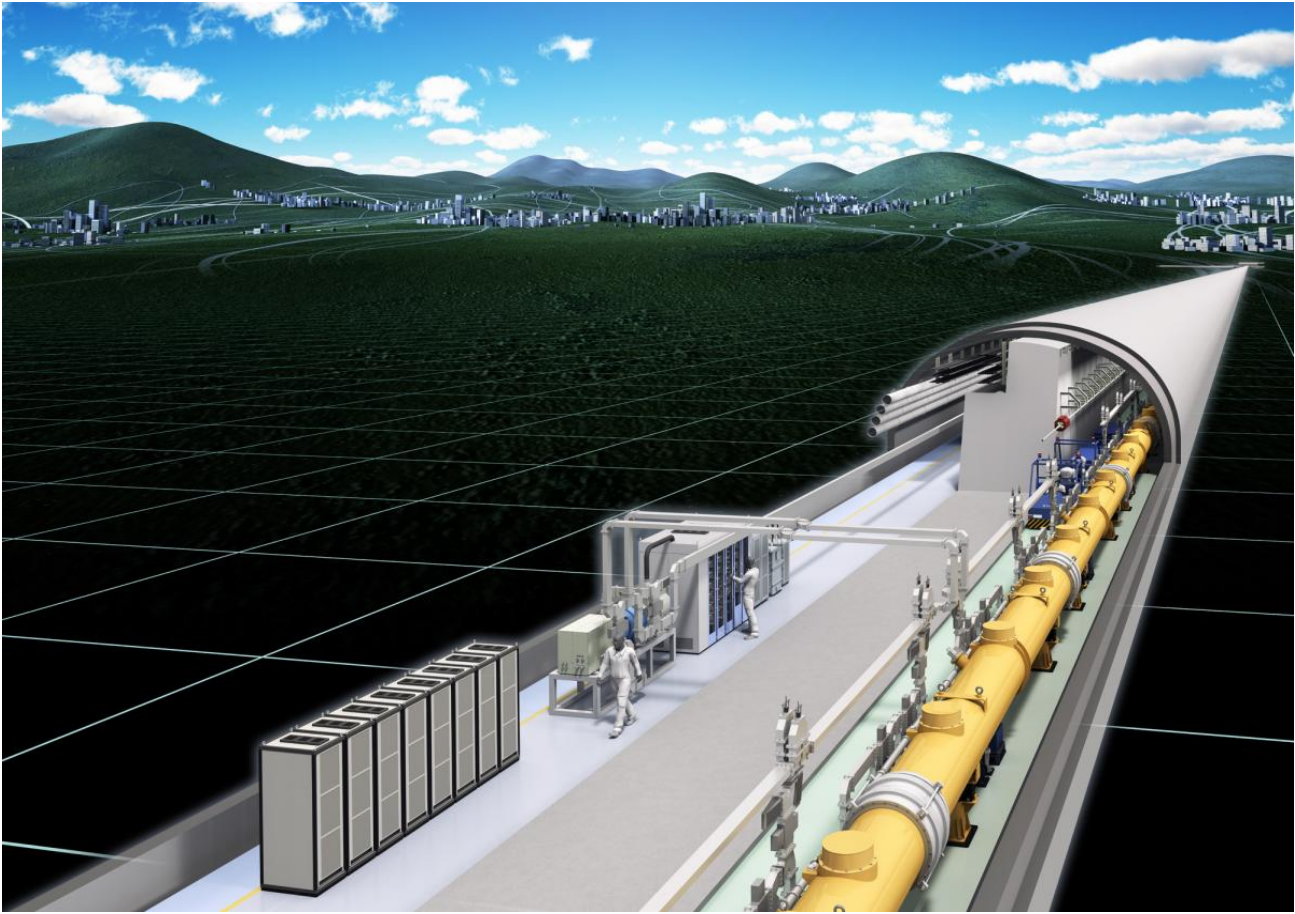

**Figure 4. Artist's rendering of the International Linear Collider.** The ILC proposed to be hosted in Japan, showing the cryomodules (in yellow on the right) of the linear accelerator that contain the superconducting radiofrequency accelerating cavities and the power units and distribution system within the tunnel. (Credit: Rey.Hori/KEK)

**Strategic Planning Processes**

For the European Strategy for Particle Physics Update that is currently underway, the scenarios proposed [49] are for facilities at CERN such as CLIC, FCC (ee, hh), the high-energy LHC and variations of them. However, the European Strategy Update must necessarily take into consideration proposals in other regions of the world. The key facilities and experiments for accelerator-based neutrino physics are in the United States and Asia, and are recognised as such in the European considerations. The outcome of the Japanese deliberations on hosting the ILC, although not known in time for the strategy discussions, must be kept as a decision point for future European plans. Similarly, progress on the Chinese CEPC will be part of the considerations.

The current U.S. high-energy physics programme is guided by the strategic plan laid out in the 2014 P5 report [50]. This report followed the year-long community-wide study, the "Snowmass" 2013 [51] organized by the Division of Particles and Fields of the American Physical Society. The P5 strategy centred around five "science drivers": the Higgs boson as a tool for discovery; the physics of neutrino mass; identification of the new physics of dark matter; understanding of cosmic acceleration; and exploration of the unknown. The P5 plan explicitly supports the LHC, LBNF/DUNE and the ILC programmes. The next "Snowmass" community study will occur during 2020–2021 for the next foreseen P5 report in 2022.

Strong U.S. participation in the LHC construction and experimental programme was made possible by an agreement co-signed by the U.S. and CERN in 1997 that was set to expire in 2017. Given the strong endorsement of the U.S.–CERN partnership by the 2014 P5 report, a new agreement was signed with CERN in May 2015, providing for cooperation and "reciprocity". U.S. scientists will continue to participate in the LHC and high-luminosity LHC programmes. The agreement also calls for CERN's direct participation in the U.S. particle physics programme, particularly in neutrino physics.

In Asia there is no equivalent process to compare with either the European Strategy Update or the U.S. P5 process. However, in Japan a roadmap for particle physics was updated in 2017 [52] by the Committee on Future Projects in High Energy Physics of the Japan Association of High Energy Physicists (JAHEP). The strategy makes a compelling case for construction of the 250 GeV ILC in Japan without delay to investigate detailed properties of the Higgs boson.

JAHEP also encouraged realization of the Hyper-K neutrino and proton decay project with international participation, and upgrades to the proton beam power at J-PARC to achieve the required neutrino beam intensity. Understanding CP violation in the lepton sector is seen as a study of major significance for particle physics.

**Particle Physics as a global enterprise**

Particle physics is globally connected. International co-operation through collaboration, balanced by competition has driven the field to astounding success while straddling different political and funding systems. Innovative, ever more powerful facilities and experiments have been the hallmark of the field.

With the success of the LHC, particle physics has demonstrated the building and exploitation of very large facilities with multinational contributions. The field is clear about the need for a dedicated electron–positron Higgs factory as the critical next generation global facility. The ILC will be such a facility and perhaps the next major facility to reach approval. Unlike the LHC built at a well-established high-energy physics laboratory, the ILC will depend upon the willingness and capacity of a single government to launch a new international laboratory with negotiated international commitments. At the most recent ICFA meeting, Japan's MEXT and National Diet officials reaffirmed strong interest in realizing the ILC in Japan and plan to continue international negotiations [53]. It remains to be seen how major laboratories and nations around the world can adapt to, and embrace, this expanded level of cooperation.

Longer term energy frontier options are being pursued with vigour, although which option to choose is less clear. CERN and the United States have driven the energy frontier for decades. CERN is now without a cohort-competitor in its capacity to continue the role in the near-term, as the United States focuses on the neutrino and precision frontiers. However, a new round of community study beginning in the United States could bring new ideas and aspirations. The impact of major programmes in Asia, in particular China, towards the future energy frontier could also be very significant, although hurdles have to be overcome in international relations to embrace a major international facility hosted in China. At present, however, it is palpable that the current European Strategy for Particle Physics Update will have a critical role in setting future directions for the field globally.

Breakthroughs in enabling accelerator technologies could be game-changers for the longer-term future. The technologies developed in the pursuit of elementary particles have found transformative applications in other disciplines — from life sciences to designer materials — with great societal benefits. Future critical challenges for the field of particle physics to meet its scientific demands and advance frontiers of knowledge would include overcoming potential geo-political barriers and resource limitations. We urge a global optimisation of international resources to ensure continued development of the extraordinarily successful field of high-energy particle physics.

**Acknowledgments:**

P.C.B. is supported by Fermi National Accelerator Laboratory (FNAL/Fermilab), which is managed by Fermi Research Alliance, LLC (FRA), under the contract number DE-AC02-07CH11359 with the U.S. Department of Energy. G.N.T. is supported by the Australian Research Council and the University of Melbourne.